\documentclass[preprint2]{emulateapj}

\slugcomment{version 2010 Jan 31}

\shorttitle{Gamma Ray Burst prompt emission}
\shortauthors{F. Massaro \& J. E. Grindlay 2010}

\begin{document}

\title{Spectral evolution of long Gamma Ray Burst prompt emission:\\
electrostatic acceleration and adiabatic expansion}%@title
\author{F. Massaro\altaffilmark{1}\& J. E. Grindlay\altaffilmark{1}}
\affil{Harvard - Smithsonian Center for Astrophysics, 60 Garden Street, Cambridge, MA 02138}

\begin{abstract} %@abs
Despite the great variation in the light curves of Gamma Ray Burst (GRB) prompt emission,
their spectral energy distribution is generally curved and broadly peaked.
In particular, their spectral evolution is well described by the
hardness-intensity correlation during a single pulse decay phase, when
the SED peak height $S_p$ decreases as its peak energy $E_p$ decreases. 
We propose an acceleration scenario, based on electrostatic
acceleration, to interpret the $E_p$ distribution peak at $\sim$ 0.25 MeV. 
We show that during the decay phase of individual pulses in the long GRB light
curve, the adiabatic expansion losses likely dominate the synchrotron cooling effects.
The energy loss as due to adiabatic expansion can also be used to describe the spectral evolution 
observed during their decay phase.
The spectral evolution predicted by our scenario is consistent
with that observed in single pulses of long BATSE GRBs.  
\end{abstract}


\section{Introduction}
The acceleration mechanisms and the radiative processes 
underlying the prompt emission in long GRBs are still unclear.
Their large energies released  on short timescales likely 
requires that the radiation is produced in
a highly relativistic jet (e.g. M\'esz\'aros 2002).

We may consider that a GRB consists of fundamental units of emission,
or {\it pulses}, in the light curve (e.g. Norris et al. 1996, Stern \& Svensson 1996).
The pulse structures generally show
a sharp rise and a slower decay phase to the background flux threshold.
However, GRBs exhibit a wide variety of light curves, both in shape and in duration, 
and the combination of many such pulses could create the
observed diversity and complexity of light curves (Fishman et al. 1994). 

GRB spectral energy distributions (SEDs) are generally curved and 
broadly peaked. They are usually well 
described by the Band function (Band et al. 1993) or, as shown more recently,
by the simpler (and physically motivated) log-parabolic model 
(Massaro et al. 2010, hereinafter M10).
Kaneko et al. (2006, 2008) investigated the time resolved spectral behavior of the GRBs 
in the BATSE catalog (Paciesas et al. 1999) and
showed that the distribution of their SED energy peaks, $E_p$,
is symmetric around $\sim$ 0.25 MeV (Goldstein et al. 2010).

As reported in the analysis of 
Ryde \& Svensson (2002), the GRB spectral evolution during the decay phase of individual pulses
can be described by the hardness-intensity correlation (HIC) 
between the time resolved SED peak height, $S_p$ 
(defined as $\nu F_{\nu}$ and proportional to the total flux measured
at $E_p$), and the peak energy $E_p$  in the form of a power-law:
$S_p \propto E_p^{\,\eta}$,
where the distribution of the $\eta$ parameter peaks  at value $\sim$ 1.7 (Borgonovo \& Ryde 2001).

Subsequently, Ryde \& Petrosian (2002) showed that 
a powerlaw form (i.e. $S_p \propto E_p^{\,\eta}$) can be reproduced through
kinematic effects when applied to a spherical
shell expanding at extreme relativistic velocity. 
The {\it curvature} of a relativistic shell would make the photons emitted
off the line of sight delayed and affected by a varying
Doppler boost as a result of the increasing angle at which
the photons are emitted with respect to the observer. 
They show that these so called {\it curvature effects}, characterized 
by a time scale $\tau_{ang}$,
display a similar trend to that of the hardness-intensity correlation 
(HIC) observed in the GRB spectral evolution,
with the parameter $\eta$ =2 then expected.
However, they also argued that an intrinsic correlation between these two spectral 
parameters in the GRB prompt emission could affect the observed HIC 
(see also Kocevski et al. 2003). These curvature effects are
dependent on the radius $R$ of the emitting shell, and are likely to
be negligible if $R$ $\leq$ 10$^{13}$ cm.

Motivated by these observations, 
we propose an acceleration scenario to explain the observed $E_p$ distribution
of individual pulses in long GRBs. 
We also show that losses for adiabatic expansion play an important role during the decay phase
and could be more relevant than synchrotron radiative cooling.
Finally, we argue that the observed spectral behavior of long pulses, during the
GRB decay phase of the light curve, can 
be described taking into account the energy loss for adiabatic expansion.

For our analysis, we use cgs units and we assume a 
flat cosmology with $H_0=72$ km s$^{-1}$ Mpc$^{-1}$,
$\Omega_{M}=0.26$ and $\Omega_{\Lambda}=0.74$ (Dunkley et al. 2009).
Unless stated otherwise,
primed quantities refer to the observer reference frame while unprimed quantities refer to the GRB frame.

\section{Electrostatic acceleration mechanism}
We propose a particle acceleration scenario 
to explain the $E_p$ distribution around the observed value of $\sim$ 0.25 MeV.
We assume that the acceleration mechanisms occurring during the GRB
prompt emission are a combination of 
systematic acceleration, responsible for the energy peak position 
of the accelerated particle energy distribution (PED),
and stochastic acceleration, which accounts for the 
broadening of the PED around its peak (M10).

As proposed by Cavaliere \& D'Elia (2002) for Blazar jets, GRBs could be powered by 
the Blandford \& Znajek mechanism (1977) for the extraction of rotational
energy from a spinning Black Hole (BH) via the Poynting flux associated
with the surrounding magnetosphere.
In these magnetospheres, the electric fields parallel to magnetic fields can accelerate 
charged particles. They can arise, for example, as a result of magnetic field 
reconnection in current sheets or MHD jet instabilities (e.g. Litvinenko 1996, Medvedev \& Loeb 1999).

The force free condition ${\underline E} \times {\underline B} = 0$ governing these magnetospheres
breaks down when the electric field $E \leq B$.
In particular, electric fields are electrodynamically screened out at distances that exceed 
the Debye length, $d$, that for a pair plasma is defined as:
\begin{equation}
d = \frac{c}{\omega_p} = \left(\frac{\gamma~m_e~c^2}{4\pi~e^2~n}\right)^{1/2} = 5.30 \times 10^5 \left(\frac{\gamma}{n}\right)^{1/2}~cm~,
\end{equation}
where $\omega_p$ is the plasma frequency, $\gamma$ is the electron Lorentz factor, 
$m_e$ is the electron mass, $e$ its electric charge, $c$ the 
speed of light, $B$ the magnetic field and $n$ the plasma density.
Electric fields parallel to magnetic fields accelerate charged particles and
consequently, the particle energy gain for each acceleration step can be written as:
\begin{equation}
\gamma~m_e~c^2 \simeq e~B~d. 
\end{equation}
Substituting $d$ from Equation (1), we obtain an expression for the Lorentz factor
of the accelerated particle:
\begin{equation}
\gamma = \frac{1}{4~\pi~m_e~c^2} \left(\frac{B^2}{n}\right) = 9.77 \times 10^4 \left(\frac{B^2}{n}\right).
\end{equation}
We note that the above expression is similar to the assumption that the 
electron energy density $u_e \sim n~\gamma~m~c^2$ is twice the 
magnetic energy density $u_B = B^2/8\pi$, close to the equipartition condition.

With the above acceleration scenario, for an emitting region 
with particle density n $\sim$ 5$\times$10$^8$ cm$^{-3}$,
a magnetic field $B$ $\sim$ 10$^4$ G and a beaming factor 
$\delta \sim$100, all of which are typical values 
for GRB models (e.g. Zhang \& M\'esz\'aros et al. 2002), 
the synchrotron energy peak $E'_p$ is $\sim$ 0.3 MeV, in agreement 
with the observed $E'_p$ distribution.
We argue that the variance of the $E'_p$ distribution can be due to the dispersion of the 
other parameters and their intrinsic variations during the burst.
The Poynting flux in the current sheet provides its 
luminosity, which can be estimated as 
$L$ = $c B^2 /2\pi \times~(l_{cs}~w)$, 
where $w$ and $l_{cs}$ are the current sheet width 
and its length, respectively (Litvinenko 1999).

The typical observed isotropic luminosity of a GRB is $L'_{iso}$ $\sim$10$^{52}$ erg~s$^{-1}$ 
so the intrinsic equivalent value, rescaled for a beaming factor of 100, 
is $L_{int}$ = $L'_{iso}$/$\delta^4$ $\sim$ 10$^{44}$ erg~s$^{-1}$.
Assuming $l_{cs}$ $\sim$ $w$ $\sim$ 10$^{13}$ cm, as derived from the GRB variability timescale (i.e. $\sim$ 0.1 s),
the Poynting flux in a single current sheet is  $L\sim$ 10$^{44}$ erg~s$^{-1}$,
the same order of magnitude of the GRB intrinsic luminosity.

\section{Particle energy losses}
A simple scenario to describe single pulses in long GRB light curves
assumes an impulsive heating of particles and a subsequent
cooling and emission. The rise phase of pulses is
attributed to particle acceleration energizing the emitting region
while the decay phase reflects the particle energy losses.
In the following, we show that adiabatic expansion is the 
main process responsible for the particle energy losses 
during the decay phase of single pulses. This is also supported by  
the fact that the synchrotron cooling time appears too short to account for the decay of GRB pulses. 
In addition, the observational evidence that 
GRB SEDs are curved (e.g. log parabolic) and not 
the superposition of two power laws (e.g. Band function) is a strong indication that 
stochastic acceleration occurs during the prompt emission (M10). 
This suggests  that both systematic acceleration (e.g. due to 
electric fields) and stochastic acceleration mechanisms 
(e.g. due to turbulence) balance the synchrotron radiative losses.

We neglect the radiative losses from inverse Compton emission,
since GRB prompt emission does not appear to be dominated by the high energy $\gamma$-ray component 
(i.e. $\geq$ 100 MeV, Abdo et al. 2009).
%appears to be rare, as suggested by the lack of $\gamma$-ray photons
%in the LAT energy range for the majority of the GRBs 
%detected (Abdo et al. 2009). 
%Time delays also suggest that such high energy emission may 
%arise in the afterglow emission region, which is not considered in our model.

The adiabatic expansion of the emitting region can be described by  
a self similar model in which the temporal evolution of the 
radius and consequently the density, can be expressed as:  
\begin{equation}
R(t) = R_0~\left(\frac{t}{t_0}\right)^q~~,~~~n(t) = n_0~\left(\frac{t}{t_0}\right)^{-3q},
\end{equation}
where $q$ is the expansion index and is positive (i.e. $0< q\leq1$), and $t_0$ is the reference time.
The rate of expansion of the emitting region is defined by the relation:
\begin{equation}
\dot{R} = \frac{dR}{dt} = q~\frac{R_0}{t_0}~\left(\frac{t}{t_0}\right)^{q-1},
\end{equation}
and the adiabatic expansion losses of a single particle can be written as:
\begin{equation}
\dot{\gamma}_{ad} = \frac{d\gamma}{dt} = -\frac{\dot{R}}{R(t)}~\gamma~,
\end{equation}
for which the analytical solution is:
\begin{equation}
\gamma(t) = \gamma_0~\left(\frac{t}{t_0}\right)^{-q}.
\end{equation}
Assuming conservation of the magnetic flux, we can express the 
temporal evolution of its value as:
\begin{equation}
B(t) = B_0~\left(\frac{R_0}{R}\right)^{2} = B_0~\left(\frac{t}{t_0}\right)^{-2q},
\end{equation}
where $B_0$ is the initial value at time $t_0$.
We note that by replacing the relation for the magnetic flux 
conservation (Equation (8)) and for the temporal evolution of the 
particle density (Equation (4) in Equation (3), 
we obtain the same temporal evolution for the particle energy as derived in Equation (7).
Consequently, the assumption of the conservation of the magnetic 
flux is in agreement with the electrostatic acceleration scenario.  

The ratio $\rho$ between the energy losses due to synchrotron 
radiative cooling and adiabatic expansion is:
\begin{equation}
\rho = \frac{\left(dE/dt\right)_{syn}}{\left(dE/dt\right)_{ad}} = \frac{\sigma_T~B_0^2~t_0}{6\,\pi~q~m_e~c} ~\gamma ~ \left(\frac{t}{t_0}\right)^{1-4q},
\end{equation}
where $B_0$ is the magnetic field evaluated at the initial 
time $t_0$, and $\sigma_T$ is the Thomson cross section.

By substituting Equation (3) into Equation (9), the ratio of energy losses 
$\rho$ can be also written as:
\begin{equation}
\rho = 1.27 \times 10^{-4} ~\frac{t_0}{q}~\frac{B_0^4}{n_0} ~ \left(\frac{t}{t_0}\right)^{1-5q}.
\end{equation}
We assume $q > 1/5$, as in the case of constant expansion rate (i.e. $q = 1$)
or expansion with constant energy (i.e. $q=2/5$, Sedov phase condition).
Then, for typical values of the prompt emission region, during ther decay phase,
$B_0 \simeq 10^3$ G and $n \simeq 10^7$ cm$^{-3}$,
the synchrotron radiative losses are dominated by those of the adiabatic expansion (i.e. $\rho \leq 1$)
after a time interval $t_*$ of $\sim$ 1 s.
We also note that for a more compact source with  
$B_0$ $\simeq$ 10$^4$ G and $n \simeq 10^8$cm $^{-3}$,
$t_*$ is $\sim$ 10 s.

Finally, we remark that the synchrotron cooling time is dependent on 
the particle energy, with  
$\tau_{syn} = 5.16 \times 10^8 \gamma^{-1} B^{-2}\rm{s}$. In contrast,   
the energy loss time due to adiabatic expansion, 
$\tau_{ad} = \gamma/\dot{\gamma}_{ad} = R(t)/\dot{R}$ (see Equation (6),
is energy independent and so does not affect the shape of the 
particle energy distribution during the decay phase.
Therefore, if the synchrotron cooling is the main energy 
loss mechanism, we expect a drastic change in the curvature 
of the observed spectra, making them narrower with time, 
while energy losses for adiabatic expansion should preserve the 
spectral curvature (e.g. Massaro et al. 2006).

\section{Spectral evolution of pulse decay phase}
To interpret the hardness-intensity correlation we note that in a synchrotron scenario, 
the intrinsic energy peak of the spectral energy distribution 
depends on the magnetic field and the electron energy as $E_p$
$\sim$ $\gamma^2~B$, 
while the maximum of the SED evaluated at this energy is $S_p$ $\sim$ $N~\gamma^2~B^2$,
where $N$ is the total number of emitting particles (e.g. Massaro et al. 2008).
Applying the corrections due to the adiabatic expansion losses (i.e. Eq. 7 and Eq. 8) and
assuming the number of emitting particles is (approximately) constant, 
we find that the intrinsic parameters $E_p$ and $S_p$ have temporal dependence: 
\begin{equation}
E_p \propto \left(\frac{t}{t_0}\right)^{-4q}~~,~~~S_p \propto \left(\frac{t}{t_0}\right)^{-6q}.
\end{equation}
This gives the expected intrinsic relation between the two SED parameters 
$E_p$ and $S_p$ as a power-law: $S_p \propto (E_p)^{3/2}$, 
independent of the value of the expansion index $q$. 
The relativistic corrections (due to the relativistic beaming or curvature effects, e.g. Ryde \& Petrosian 2002)
do not affect the intrinsic correlation between $E_p$ and $S_p$. 
Therefore, the expected observed power-law index is still $\sim$ 1.5, and thus near the 
peak of the $\eta$ distribution estimated for the pulse decay phase of long GRBs.
In addition, assuming that the size of the emitting region 
is $\sim$10$^{13}$ cm, the curvature effects are negligible,
because their timescale is too short $\tau_{ang} \sim 10^{-2}$ s to explain the pulse decay in long GRBs.
Thus, we describe the decay phase in the GRB prompt emission
has energy loss dominated by adiabatic expansion, 
assuming that the acceleration energy gain, via systematic 
and stochastic acceleration, balances the 
synchrotron radiative cooling..

We note that the PED in the form of a log-parabolic function 
is a good approximation for the solution of 
the kinetic equation for the particles when considering terms 
taking into account systematic and stochastic acceleration as well as
including synchrotron radiative losses and adiabatic expansion
(Kardashev 1962, Tramacere et al. 2009, Paggi et al. 2009, M10).
%-----------------------------------------------------------------------------------------------------------------------
\begin{figure}[!htp]
\includegraphics[height=6.cm,width=8.5cm,angle=0]{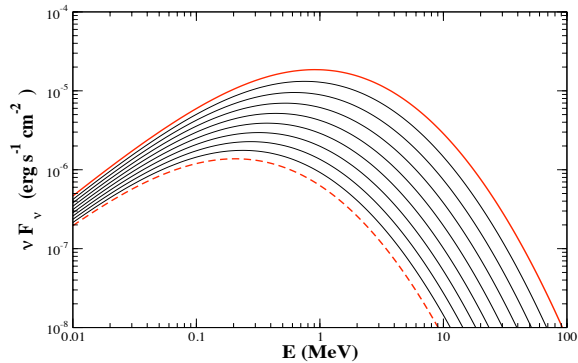}
\caption{The spectral evolution (from the thick red line to the dashed one)
of the decay phase for an adiabatically expanding synchrotron shell
based on the assumption described in Sect. 3.
Each SED is evaluated every 0.5 s for a decay phase of 6 s. 
It is evident that the shape of the SED is preserved and the curvature does not change.
In our calculations we set $B_0$ = 10$^3$ G, $n_0$
 = 10$^7$ cm$^{-3}$, $\gamma_p$ is 2$\times$10$^4$, $\delta$ = 50 and $q$ = 0.8. 
The PED spectral curvature is $r$ = 5.0 and the initial volume to 10$^{39}$ cm$^3$.}
\end{figure}
%-----------------------------------------------------------------------------------------------------------------------

We calculate the synchrotron emission of an adiabatically expanding spherical region
assuming an emitting PED with a log-parabolic shape:
$N\left({\gamma}\right)= N_0\, (\gamma/\gamma_p)^{-2-r\log{\left(\gamma/\gamma_p\right)}}.$
where $\gamma_p$ (i.e. the peak of $\gamma^2 \times N(\gamma)$) is $\langle{\gamma^2}\rangle^{1/2}$ , 
$r$ is PED curvature and $n \propto N_0$ is the density (see M10). 
Then, as shown in Figure 1, the spectral curvature of the synchrotron SED is constant during the pulse decay phase.
%-----------------------------------------------------------------------------------------------------------------------
\begin{figure}[!htp]
\includegraphics[height=3.2cm,width=8.5cm,angle=0]{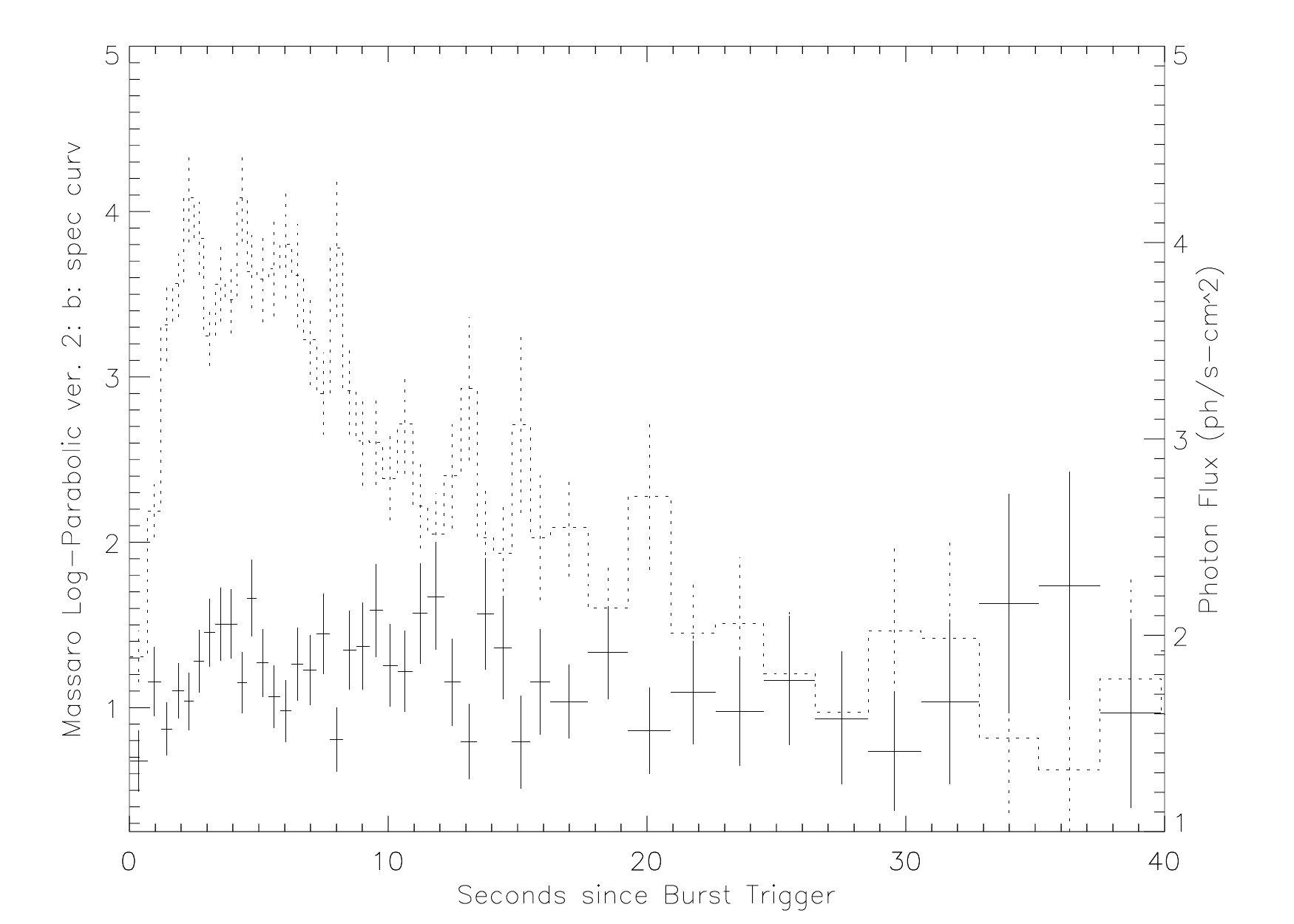}
\includegraphics[height=3.2cm,width=8.5cm,angle=0]{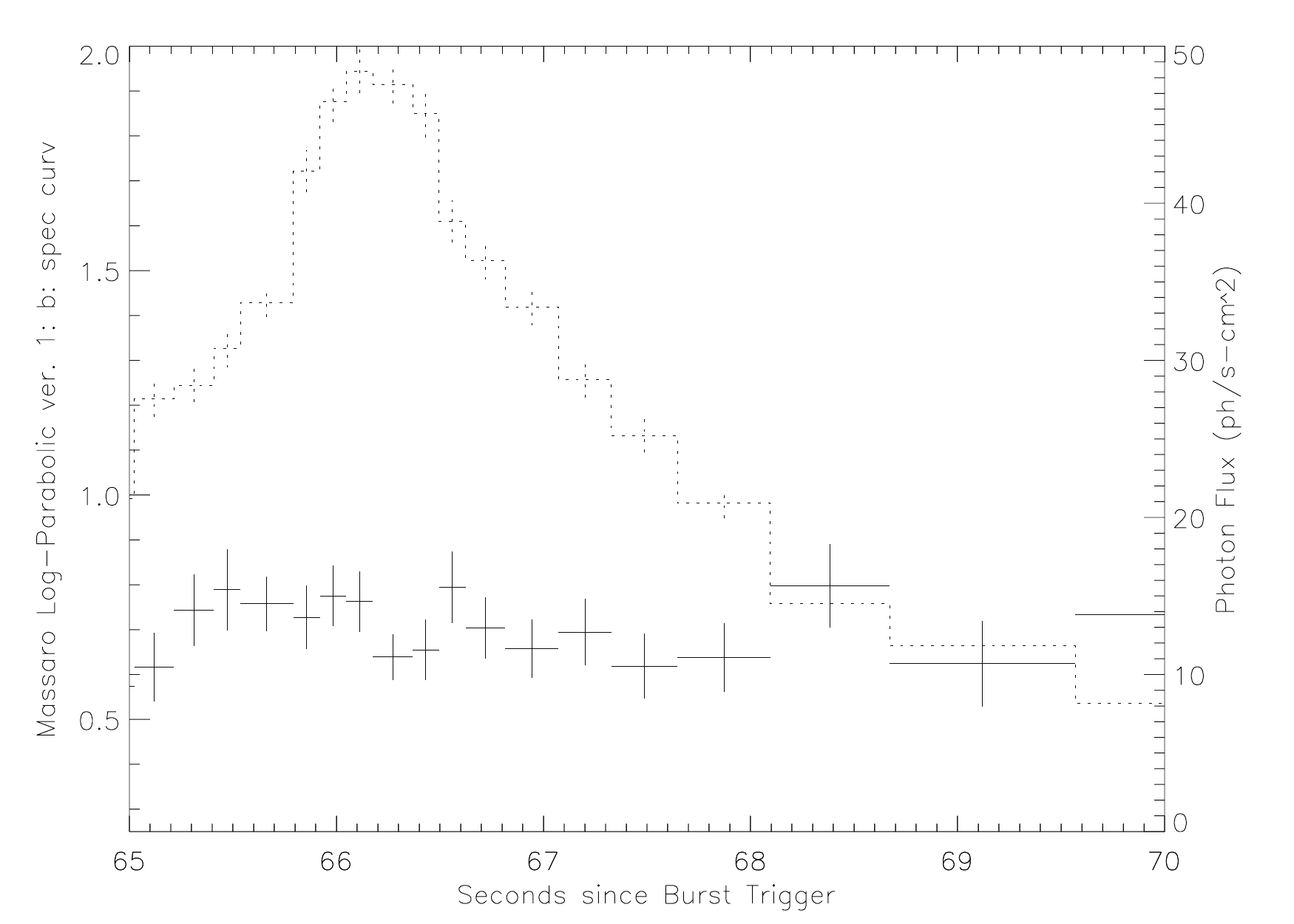}
\caption{The spectral curvature, measured during the whole light curve
of the long, single pulses in GRB 941026 (top panel) and GRB 950818 (lower panel), 
is seen to be relatively constant. 
The time resolved spectral analysis has been performed with the log-parabolic function (see M10).
The shape of the light curve is given by dashed line; right hand flux scale).
The background BATSE light curve (25 - 55 keV) is also shown (dashed line).}
\end{figure}
%-----------------------------------------------------------------------------------------------------------------------

In Figure 2, we show the time resolved spectral analysis of the decay phases for two long, single pulse BATSE GRBs,
to compare our model with the observed spectral behavior.

We adopted the log-parabolic function (i.e. $F(E) = S_p\,/E^2\, (E/E_p)^{-b\,log(E/E_p)}$, see M10) to describe 
the time resolved SED and to measure the spectral curvature $b$.  

In Figure 2, we show that, the examples of GRB 941026 and GRB 950818 do not show significant variation of their curvature $b$
during their pulse decay phases, in agreement with the scenario dominated by adiabatic expansion losses.

The detail of the spectral behavior observed during the decay phase (i.e. between 66 and 70 s) of GRB 950818 is also shown in Figure 3.
The curvature is not drastically varying over the whole burst, in agreement with our scenario (Figure 1).

We also note the presence of small fluctuations in $b$, which appear to be anti-correlated 
with secondary peaks in the GRB light curve. As already pointed out by Vetere et al. (2007)
from the analysis of the BeppoSAX WFC archive, the low energy (i.e. 2 - 30 keV)
GRB light curves are characterized by peaks superposed on a slowly evolving component. 
A more detailed analysis of these fluctuations will be presented in a forthcoming paper.

\section{Conclusions}
We propose an electrostatic acceleration scenario to 
interpret the $E_p$ distribution of GRB time resolved SEDs.
We show that taking into account adiabatic expansion 
losses it is possible to describe the spectral evolution during the decay phase of individual pulses
in long GRBs.
Our model assumes that the particle energy gain is due to 
both systematic and stochastic particle acceleration, 
while the particle energy losses are due to synchrotron emission and adiabatic expansion.

Describing the systematic acceleration in terms of electric field energy gain, we derive a simple relation
for the expected particle Lorentz factor, $\gamma$ $\sim$ 10$^4$-10$^5$.
Thus, for a typical GRB magnetic field of $\sim$10$^4$ G, a plasma density 
of $\sim$ 10$^8$ cm$^{-3}$,and a beaming factor $\sim$ 50 - 100 
the expected synchrotron peak energy $E'_p$ is $\sim$ 0.3 MeV 
as found for the observed distribution (e.g. Kaneko et al. 2006).
This may explain the non uniform time resolved $E_p$ distribution
of the GRB SED peaking around a characteristic value. 

Following the assumption that systematic and stochastic acceleration
mechanisms balance the synchrotron radiative cooling, 
and the adiabatic expansion loss is the main process governing GRB
spectral evolution during the decay phase of individual pulses, we 
derive that the expected intrinsic scaling relation between the height of the
SED $S_p$ and its peak energy $E_p$ is $S_p \propto E_p^{3/2}$, 
which agrees with the observed HIC for single pulses in long GRBs.

Finally, on the basis of our assumptions, we note that the adiabatic
losses do not change the shape of the SED during the prompt
emission. We showed that this is 
consistent with the spectral behavior of the decay phase of single pulses during long GRB prompt 
emission as in the cases of GRB 941026 and 950818, for which we did not detect 
any large variation in the spectral curvature
throughout the spectral evolution.
%-----------------------------------------------------------------------------------------------------------------------
\begin{figure}[!htp]
\includegraphics[height=5.cm,width=8.5cm,angle=0]{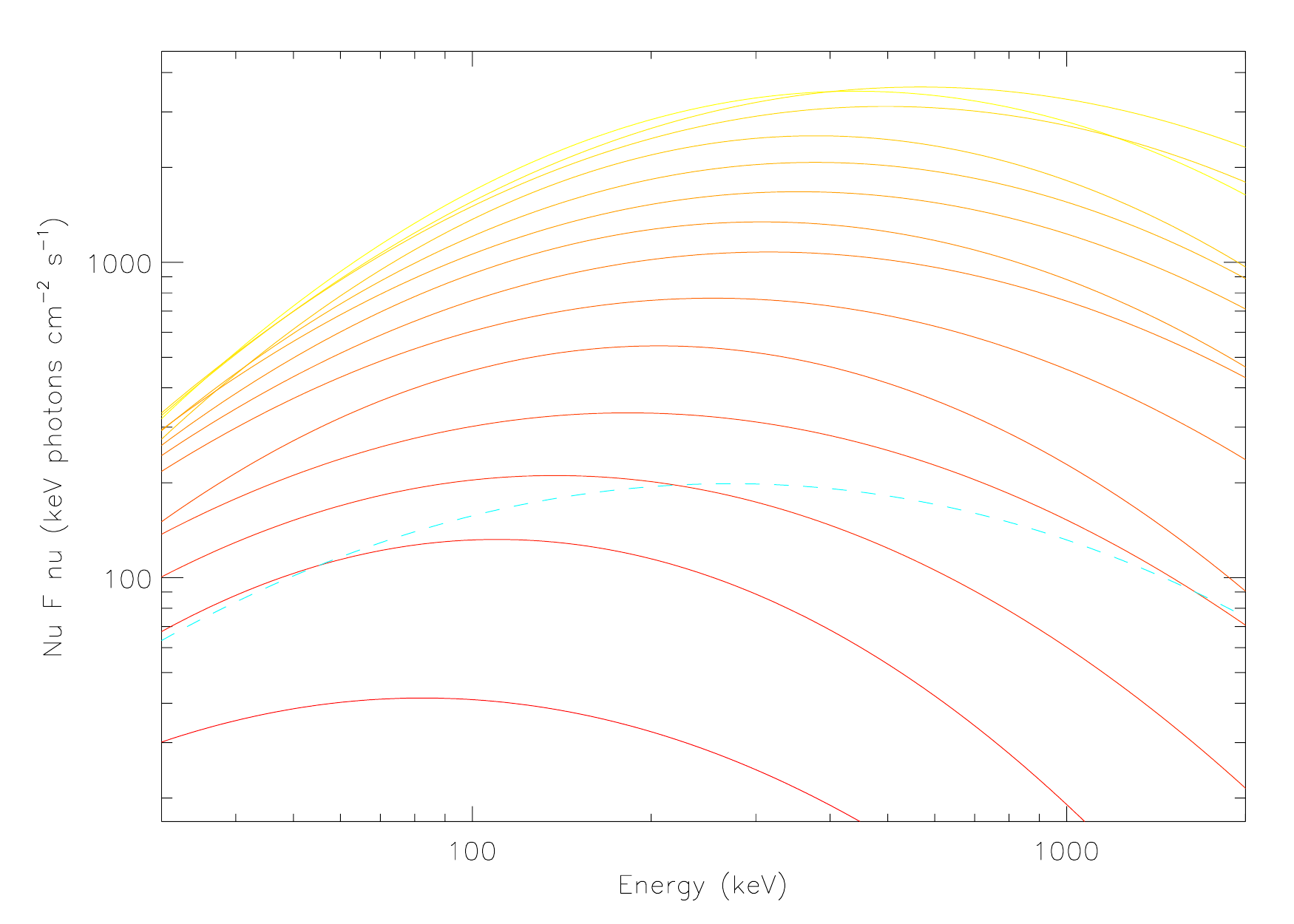}
\caption{The time resolved best fit models of the decay phase for GRB 921123 performed with the log-parabolic function. 
The time interval between each SED is 0.5 s, going from the brigthest yellow to the faintest red.
The cyan dashed line is the time integrated best-fit model.}
\end{figure}
%-----------------------------------------------------------------------------------------------------------------------

F. Massaro is grateful to R. Preece for comments to the manuscript. 
He also thanks M. Petrera, A. Cavaliere and A. Paggi
for helpful suggestions on the acceleration mechanisms.
Finally, he is grateful to M. Salvati for providing useful suggestions that helped us to improve the presentation.
F. Massaro acknowledges the Foundation BLANCEFLOR Boncompagni-Ludovisi, n'ee Bildt 
for the grant awarded him in 2010. His at SAO is also supported by NASA grant NNX10AD50G.\\

\end{document}